%% file: main_arXiv.tex
\documentclass[submission,copyright,creativecommons]{eptcs}
\providecommand{\event}{Under review for a Journal since March 2018.} %
\pdfoutput=1

\usepackage[margin=20mm,top=20mm,bottom=30mm]{geometry} %

\usepackage{graphicx}
\usepackage[utf8]{inputenc}
\usepackage{color}
\usepackage{hyperref}
\input{preamble_submission.txt}
\begin{document}

\title{InfoInternet for Education in the Global South: \\ A Study of Applications Enabled by Free Information-only Internet Access in Technologically Disadvantaged Areas \\ (authors' version)} 

\def\titlerunning{InfoInternet for Education in the Global South (authors' version)}
\def\authorrunning{
J. Johansen, C. Johansen, J. Noll
}

\forSubmissionOnly{
\author{Anonymous Authors for Submission only
}
}
\forFinal{
\author{Johanna Johansen
\institute{Department of Informatics, University of Oslo}
\email{johanna@johansenresearch.info}
\and Christian Johansen 
\institute{Department of Technology Systems, University of Oslo}
\email{cristi@ifi.uio.no}
\and Josef Noll
\institute{Department of Technology Systems, University of Oslo}
\email{josef@jnoll.net}
}
}

\maketitle

\begin{abstract}
\input{abstract_submission.txt}

\vspace{1ex}
\end{abstract}

\footnotetext{\textbf{Acknowledgements:} This work was partially supported by the project \href{https://DigI.BasicInternet.org}{DigI} -- Non-discriminating access for Digital Inclusion, part of the \href{https://www.norad.no/visjon2030}{Visjon 2030} program funded by the \href{http://www.forskningsradet.no}{Norwegian Research Council}.
}

\input{introduction_submission.txt}
\input{infointernet_submission.txt}

\input{methodology_submission.txt}

\input{related_work_submission.txt}

\input{users_submission.txt}

\input{education_submission.txt}

\input{examples_submission.txt}

\input{conclusion_submission.txt}

\input{main_arXiv.bbl}
\end{document}

%% file: preamble_submission.txt
\newcommand{\cj}[1]{} %
\newcommand{\cjLONG}[1]{}%
\newcommand{\jnLONG}[1]{} %
\newcommand{\jj}[1]{} %
\newcommand{\jjTODO}[1]{} %
\newcommand{\jjTODOnext}[1]{}
\newcommand{\mynotes}[1]{} %
\newcommand{\forlong}[1]{} %
\newcommand{\forFinal}[1]{#1} %
\newcommand{\forSubmissionOnly}[1]{} %

%% file: abstract_submission.txt
This paper summarises our work on studying educational applications enabled by the introduction of a new information layer called InfoInternet.
This is an initiative to facilitate affordable access to internet based information in communities with network scarcity or economic problems from the Global South. 
InfoInternet develops both networking solutions as well as business and social models, together with actors like mobile operators and government organisations.
In this paper we identify and describe characteristics of educational applications, their specific users, and learning environment.
We are interested in applications that make the adoption of Internet faster, cheaper, and wider in such communities.
When developing new applications (or adopting existing ones) for such constrained environments, this work acts as initial guidelines prior to field studies.

%% file: introduction_submission.txt
\section{Introduction}\label{sec_intro}

We use InfoInternet as an umbrella term for the activities of the Basic Internet Foundation\footnote{\url{http://basicinternet.org/} (Not to be confused with the Free Basics of Facebook.)},
which has the general goal to provide free and affordable access to basic information in areas with no or scarce Internet infrastructure. A multi-tenant deployment plan is being executed\footnote{DigI project ``Non-discriminating Access for Digital Inclusion'': \forSubmissionOnly{\url{http://its-wiki.no/wiki/DigI:Home}}\forFinal{\url{DigI.BasicInternet.org}}}, 
through pilots in rural areas in Tanzania and DRC (Democratic Republic of the Congo).
InfoInternet (see Section~\ref{sec_infointernet} for a more detailed presentation) is meant to initially help in three main social environments:
health, education, and small businesses for empowering women. Women and youth are prioritized, as the population most vulnerable in a society.
This paper focuses on the \textit{education} aspect; in particular, on studying \textit{educational applications} enabled by InfoInternet for the intended communities.

The map of Internet speed provided by Akamai\footnote{\url{https://www.akamai.com/us/en/about/our-thinking/state-of-the-Internet-report}} shows that the countries in the  Global South \cite{mitlin2013urban} have least coverage, as well as lowest speed. 
These countries also have the largest growth of population, as well as the poorest.
For children that live in poverty (ca.~20\% of world's children live in ``extreme poverty'')\footnote{\url{https://data.unicef.org/wp-content/uploads/2017/09/Ending_Extreme_Poverty_A_Focus_on_Children_Oct_2016.pdf} \\ \  \\ \ } it is often difficult to focus on learning (even when having access to schools) due to multiple traumas they have suffered. Causes for these traumas include poverty, domestic violence, parents with mental illness or under addictions, lack of home or living in refugee camps, war.
The first goal in the \textit{2030 Agenda for Sustainable Development (SDGs)}\footnote{\url{https://sustainabledevelopment.un.org/sdgs}} is the elimination of poverty, while the quality of education is number four.\footnote{\url{https://data.unicef.org/children-sustainable-development-goals/}} 
InfoInternet addresses communities having both these problems, with the belief that better education would help solve the other problems as well.
420 million people would be lifted out of poverty with a secondary education. Thus, reducing the number of poor worldwide to more than half.\footnote{Policy paper 32 / Fact sheet 44, \textit{Reducing global poverty through universal primary and secondary education}, by UNESCO Institute for Statistics (UIS) and EFA Global Education Monitoring Report, 2017, available at \url{http://uis.unesco.org/sites/default/files/documents/reducing-global-poverty-through-universal-primary-secondary-education.pdf}}

By offering free access to digital information in schools, InfoInternet aims to improve the accessibility, breadth, and quality of education in the Global South. 
Our interest in supporting ICT-based education for this vulnerable population is motivated by the fact that the \textit{participation in the digital society and the access to information} is a human right \cite{APC2016report}. 
Providing free and affordable access to information has the potential to reduce the digital divide \cite{internetsociety2017digitaldivide}, bring social change \cite{Franquesa2017SPDC}, and become a catalyst for the SDGs.

Our methodology, detailed in Section~\ref{sec_methodology}, uses Informed Grounded Theory \cite{Thornberg12GTinformed,groundingTheory2015book} to learn from existing studies about education and children in the Global South areas.
We identify the users' characteristics and needs, in Section~\ref{sec_users}, and the educational environment and how it would be affected by InfoInternet, in Section~\ref{sec_education}.
We proceed to study educational applications, in Section~\ref{sec_examples}, to extract their characteristics wrt.\ these specific users and learning environments.
The knowledge synthesised out of existing literature need to be verified and confirmed through our own studies in the field. 
\jjTODOnext{We evaluate the kind of interactions the child does with each type of application, and the kind of helping cues need to be provided by the application to achieve its educational intended purpose. \cj{Do we do this? NOT now, but in the next version we should do it, because it is simple. Maybe even the next version of the Journal, i.e. in the final version.}}
We are working with \textit{two sources of variability}: 
(I) one coming from the specific needs of disadvantaged children (like long distance to school or need to do domestic work);
(II) another coming from the special characteristics of InfoInternet, identified in Section~\ref{sec_infointernet} (like low bandwidth, slow response time, or special kind of content).

We also identify what kind of applications are likely to make the adoption of the InfoInternet 
\textit{larger} (i.e., spread easier and wider, to other areas), 
\textit{faster} (in terms of time), and 
\textit{cheaper} (in terms of resources, knowledge and training, or support from government or industry). 

%% file: infointernet_submission.txt
\section{Background on InfoInternet}\label{sec_infointernet}

The technical part of InfoInternet works with developing and deploying the networking infrastructure, whereas business and social activities are conducted to ensure its adoption. 
In this paper we start from the assumption that the network infrastructure is already put in place by InfoInternet, with the restrictions of limited and/or unstable network connections. 

\subsection{Relevant technical aspects of InfoInternet}

In this section we briefly describe the technological aspects of the InfoInternet infrastructure that are relevant for the present paper, i.e., within the educational sector, and the business model InfoInternet is based on.

\begin{figure}
\centering
  \includegraphics[width=0.99\columnwidth]{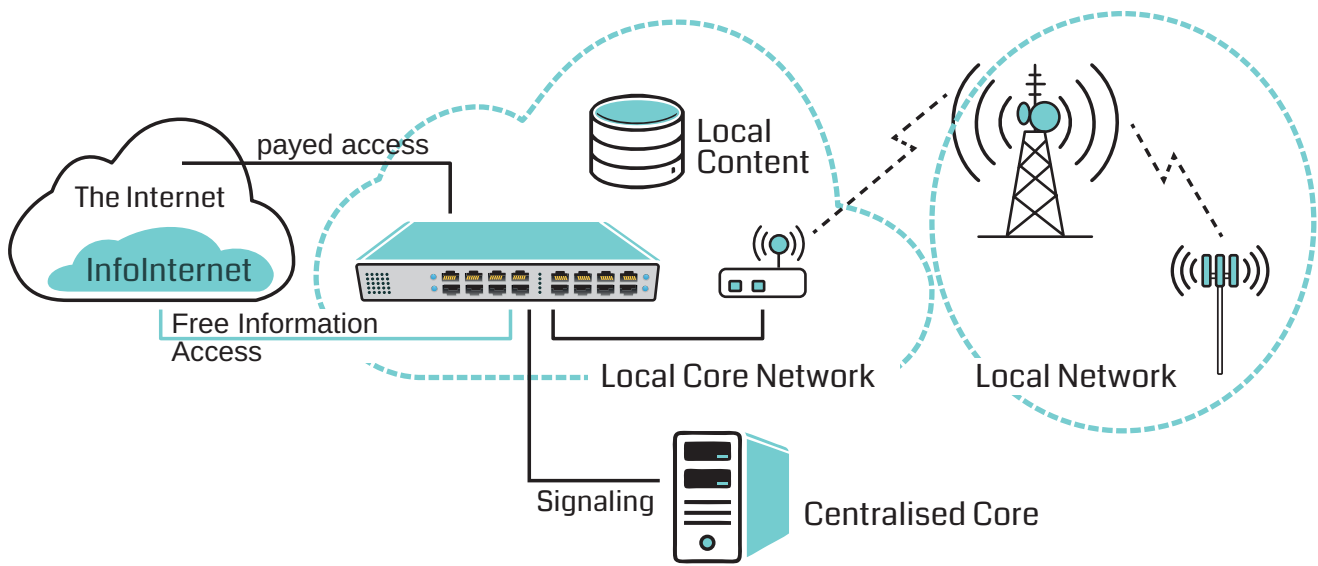}
  \caption{InfoInternet deployment architecture.}~\label{fig_BasicArchitecture}
\end{figure}

The architecture of InfoInternet, designed for low-cost local infrastructure and rapid deployment, consists of (see Fig.~\ref{fig_BasicArchitecture}):
(I) a local core network with local content, 
(II) a local network,
(III) a centralised core, 
(IV) and the backhaul network/network termination (achieved through, e.g., a radio link or a satellite connection). 
This architecture fosters
high capacity access to local content, 
free access to basic information from the Internet,
as well as paid access to other bandwidth heavy content, like video. 

The HTTP Archive\footnote{\url{http://httparchive.org/interesting.php}} shows that an average web page has doubled in size from 1.09 MB in 2012 to 3 MB in 2017, divided into scripts ca.~19\%, images ca.~60\%, while video in 2017 is accounting for ca.~24\% of the whole web size.
On a satellite link of 1 Mbps, a web page of 3 MB would load in 24 seconds blocking the capacity for all users.
Because of this, InfoInternet \cite{noll2018InfoInternet} makes the local network and local content accessible through high capacity wifi hotspots, also using caching techniques.
However, when non-local content is requested, applications need to be developed taking into consideration a low-capacity backhaul or a limited payment plan.

\textbf{The business model} suggested by InfoInternet adopts the freemium style \cite{Anderson2009Freemium,Seufert2014Freemium}, where access to basic information is free and full Internet access (e.g., to videos, games, full web-sites) is paid, e.g., through specific data plans offered by the mobile operators.
InfoInternet pilots demonstrated that the free services take ca.\,2.5\% of the bandwidth, leaving the rest for commercial operations.
This percentage can be covered  by the businesses through various schemes like tax or other state supported funding programs. Thus, InfoInternet is not proposing a diluted or in any way an inferior mobile service.

\subsection{Related projects and initiatives}

From the related projects already identified previously by InfoInternet, we mention here
Free Basics from Facebook,\footnote{\url{info.internet.org/en/story/free-basics-from-internet-org}} 
a partnership between Facebook and several companies to bring affordable access to selected web-sites.
However, their model is criticised for violating net neutrality.\footnote{\url{https://www.theguardian.com/technology/2016/feb/08/india-facebook-free-basics-net-neutrality-row}}
In contrast, non-discriminating access to basic information and net neutrality are part of InfoInternet. Moreover, InfoInternet is not dependent on specific operating systems nor apps on the users' devices.
Besides, InfoInternet can integrate with relevant
companies like 
WaveTek Nigeria LTD\footnote{WaveTek Nigeria Limited: \url{http://www.wavetek.net/}},
research programs like 
EU's Next Generation Internet\footnote{EU's Next Generation Internet (2016): \url{https://www.ngi.eu/}}, 
social initiatives like
Internet For All\footnote{\url{https://internetforall.one.org/}},
academic projects like
Gram Marg from IIT Bombay\footnote{Gram Marg project of IIT Bombay in India: \url{http://grammarg.in/}},
or business fora like
Digital Impact Alliance\footnote{\url{https://digitalimpactalliance.org}}.

\subsection{Limitations of InfoInternet}

\begin{description}
 \item[Textual information] requires a certain level of literacy from its users, e.g.,
\cite{Medhi2011TOCHI} shows that textual interfaces are unusable by first-time low-literate users, and error prone for literate but novice users. 
However, in applications for medical treatment at home or online banking, illustrations proved \cite{Medhi2011TOCHI} to be more effective and preferred by the users. 
\textit{Low resolution pictures} can be delivered by InfoInternet and thus considered for similar needs in educational applications. 

\item[Limiting streaming videos] can be an important short-coming for education, especially for ICT, since many lectures and tutorials are made available in this form. Moreover, children with different attention and reading disorders have better learning results when video alternatives exist. 
InfoInternet does not put restrictions on such type of content, providing free access to videos on the local content server as well as freemium access to any other content. 
\end{description}

\subsection{Devices relevant for InfoInternet and their usage}

InfoInternet supports all types of devices as long as they can be connected to the Internet (e.g., tablets, smartphones, PCs).

As hardware, tablets seem to be the most preferred device to be used for education. This preference is motivated by cost, adaptability and scalability but also for supporting a new kind of learning called ``here and there'', that allows for more flexibility when it comes to the physical space and time where learning can happen \cite{hassler2016tablet, martin2013here}.

%% file: methodology_submission.txt
\section{Methodology}\label{sec_methodology}

We follow an Informed Grounded Theory methodology \cite{Thornberg12GTinformed} in the constructivist style \cite{Charmaz2014GTconstrBook} to investigate the various existing applications for education that are enabled by Internet and networked communication in general. We particularly focus on applications relevant for InfoInternet and extract what exactly it means to be ``InfoInternet ready''. Our focus on the Global South areas and children living in remote areas without communication infrastructure, introduces another variability line.

As such, our method is based on the widely used Grounded Theory methods \cite{groundingTheory2015book,morse2016GTcompile,Charmaz2014GTconstrBook} to build gradually our criteria and concepts that we are looking for in an incremental fashion. Opposed to grounded theory we do not collect data ourselves, but base our work on the data collected by the previous research efforts that we cite, and rely on the conclusions that those authors drew. 

Hence, in the tradition of informed grounded theory, we study existing literature and initiatives relevant to our topic to enlarge our understanding on how access to Internet-based content will impact the quality of education in marginalised regions. 
We identify the needs and the context of use specific to undeserved communities wrt.\ information technology and Internet infrastructures. We look at the research done on digital inclusion 
\cite{Madon2009DigitalInclusion,Sanchez2010IDC} 
in rural schools around the world and use it as starting point. We also consider how access to Internet might influence the education in these communities. 
We use this research to understand better the needs and the dynamics of such communities and their influence on the adoption and use of InfoInternet in schools.

The work for this paper did not interact directly with users.
However, we used results from studies done directly both with children and students at university level in Global South. These results were mostly not peer-reviewed, but either available on the websites of the respective projects/pilots,\footnote{\url{http://its-wiki.no/wiki/Research_at_the_University_of_Lisala}}$^{,}$\footnote{DigI project ``Non-discriminating Access for Digital Inclusion'': \forSubmissionOnly{\url{http://its-wiki.no/wiki/DigI:Home}}\forFinal{\url{DigI.BasicInternet.org}}} or received directly from the field scientists that performed them. 
Since its start in 2010, the Basic Internet Foundation has run a series of pilots in Africa and Germany (now planning pilots in India), of which one established in 2011 basic Internet access for the region and University of Lisala in DRC. Another pilot\footnote{Together with Caritas Kinderdorf: \url{http://its-wiki.no/wiki/Kinderdorf:Media}} had about 60 children with specific social and psychological needs (e.g., due to violent family backgrounds) and up to 40 grown-ups looking after them, in a special community in the small town Bottrop, Germany.

%% file: related_work_submission.txt
\section{Relevant work}\label{sec_related}

\begin{description}

\item[\textbf{ICT for education}] is essential for allowing children to become contributors to the creation of technology, instead of only consumers \cite{Unnikrishnan2016OfElephants,crompton2017TheUse}.
Free access to information through InfoInternet would inspire and encourage various initiatives to reach the Global South schools, s.a.: Code Club\footnote{\url{https://www.codeclub.org.uk/}} \cite{smith2014codeClub}, Hour of Code\footnote{\url{https://hourofcode.com}}, Teach Kids Code\footnote{``L\ae{}r Kidsa Koding'' (in Norway): \url{https://kidsakoder.no}} \cite{Corneliussen16LarKoding}, or intuitive programming languages like Google's Blockly \cite{trower2015creating,weintrop2017comparing} or MIT's Scratch \cite{resnick2009scratch,maloney2010scratch} and other educational programming languages \cite{weintrop2017comparing,armoni2015scratch}.

\item[\textbf{ICT for Development (ICT4D)}] is a large field \cite{Heeks2008ict4d,Avgerou2008ict4d,unwin2009ict4d,Heeks2002ict4dBADs}, and the work in this paper fits particularly in the education part \cite{selwyn2013education,selwyn2016education}, which ICT influences both positively \cite{Pegrum14mobilEdu} and negatively \cite{selwyn2013distrustingICT4Dedu}.
InfoInternet participates in all three aspects of the ``Cube Framework'' \cite{hilbert2012toward}, i.e.: building a communication infrastructure, 
working in three socioeconomic areas, and proposing policies together with local governments and companies.

Earlier ICT4D efforts focused on the expansion of the telecommunication infrastructure. This is still needed in some parts of the global south where InfoInternet is active. However, ICT4D 2.0 encourages to use the power of low-cost devices such as mobile phones or tablets, and web 2.0 technologies to help improve access to educational content, which is what InfoInternet sees as devices, moving away from the traditional personal computers.
The use of mobile devices has been shown to be effective in improving the quality and equality of education in underdeveloped countries \cite{valk2010using,kim2012comparative} due to their increased accessibility and flexibility.

\item[Misinformation] is characterised as ``unintentional mistakes'', as opposed to \textit{disinformation} which is ``false information spread deliberately to deceive'' \cite{Mintz2002misinformation}. 
InfoInternet would be used for obtaining information by the child through ``surfing'' the web. This is vulnerable to disinformation and misinformation.
Disinformation is most often spread through social media and news channels and forums \cite{Benham2012misinformation}. 
As such, we expect \textit{misinformation} to be the main problem for InfoInternet in education \cite{Graham2003ACMmisinfo,leu2015income}.
For this the children need to be trained to detect such mistakes, like everyone else \cite{derman2010anti,Kumar2014misinfodetect,Azpiazu2017}, under the guidance of tutors or teachers.
\textit{On-line risks}, like Cyberbuling or online safety for children \cite{PinterWXRC17idc} is also a relevant aspect that needs to be considered.

\item[\textbf{Educator training.}] All school ages are expected to be affected by InfoInternet, and thus teachers for all levels need to be trained into using and making most out of new resources available through InfoInternet \cite{gordon2013beginnings,gestwicki2015home}. 
In the same school often there are notable differences between classes at the same level and even between individuals in the same class. 
Moreover, since InfoInternet will penetrate also the community and family, some resources would also contribute to the e-learning process of children at home \cite{gestwicki2015home}.

\end{description}

%% file: users_submission.txt
\section{The intended users of InfoInternet}\label{sec_users}

\subsection{Unconnected communities in the Global South}
According to the Internet Society \cite{2015InternetReport}, the majority of the Global South population does not have adequate Internet access, i.e., is slow, unreliable, and often offline; 
only ca.\,30\% having 3G coverage in 2013 and a small minority having 4G, where smartphone subscriptions in Sub-Saharan Africa were only ca.\,25\% of the population in 2016. This implies that the digital world cannot be reached by everyone fairly and thus an access gap is created between citizens who can afford a digital device and an Internet connection and those who cannot. The later represents the general user group of InfoInternet.

\textbf{Citizens} unable to access digital tools are often confined to the \textit{lower or peripheral edge of the society}, both economically and geographically. 
As a result of this inaccessibility, such groups are denied full involvement in mainstream economic, political, cultural, and social activities. This usually also implies restricted access to or exclusion from critical services such as health, education, and other public services, and therefore limited opportunities for development and welfare \cite{APC2016report}. 
The technological purpose of the InfoInternet project regards mostly the network connectivity and deployments of the infrastructure necessary to provide this. However, there are other aspects of InfoInternet like business models and PR that come into play to support the participation of our users in the digital world and even out the digital gap.

\textbf{Children} from these poor communities are \textit{the primary users regarded in this paper}. The adult part of population will only be mentioned as part of the context in which the children are growing up. Their characteristics and needs are strongly influenced by the socio-economic context they are surrounded by (we give more information about this in Section~\ref{sec_education}). 

\subsubsection{Characteristics:} 
\begin{itemize}
\item Little knowledge of what a digital device can do and how to operate it, as there is little technology accessible in these areas (a TV and an occasional smartphone)\cite{Unnikrishnan2016OfElephants}.
\item Lack support from parents, because of illiteracy or little understanding for why their children need to learn technology.
\item Irregular school attendance, e.g., due to domestic work load.
\item Have less knowledge (in school subjects) than expected for their age level in more developed countries \cite{gordon2013beginnings}.
\item Some children have psychological problems caused by a challenging home environment, which affects, e.g., their attention span, curiosity, and motivation for learning.
\end{itemize}

\subsubsection{Needs:}
\begin{itemize}
\item Access to remote and flexible learning for children that cannot attend school regularly or within the school time schedule.
\item Access to digital content translated to the child's language and relevant for their own age and individual capabilities.
\item Easy access to digital devices and Internet.
\item Guidance and help with differentiating good/appropriate from misleading/flawed information.
\item Interaction with people with the will and competence to share their ICT knowledge (friends, family).
\item Technology use should be supervised/guided by adults, so that educational uses would outweigh the entertainment ones.
\end{itemize}

\subsection{Other stakeholders}
While the main stakeholders that we consider here are children from unconnected communities, the larger scope of InfoInternet is to help all sorts of users forced to work in different contexts than the classical office space.
These are individuals that work in professional isolation, freelancers that would like to work from remote places, commuters, or researchers involved in projects meant to help these communities. 
The researchers from the Community Lab of the Faculty of Computing and Informatics (FCI), at Namibia University of Science and Technology, mention that having a mobile lab, equipped with Internet and electricity will be very useful for their research work within the community setting \cite{Winschiers2017CommunityLab}.

%% file: education_submission.txt
\section{Educational Context of InfoInternet}\label{sec_education}

\jjTODOnext{TO DO: Add also about how technology helps in ``Adaptive education''}

An effective use of ICT has three dimensions: motivation, possession, and digital skills \cite{Sanchez2010IDC}. 
\textit{Motivation} refers to the willingness of individuals to use technology and to include it in their home, work, or educational efforts \cite{kumar2017usability}.
\textit{Skills} refer to the abilities to use the technology.
\textit{Possession} is the dimension most relevant for InfoInternet, as it includes physical access to Internet and computers (or Internet-enabled devices). 
By offering free access to basic forms of information from the Internet, InfoInternet enables a baseline in the possession dimension. 
However, the other dimensions are of equal importance as they influence the adoption of InfoInternet. 
The communities need time to discover or be taught the benefits of information technologies. They need to experience the positive effects and, even more, to adapt ICT use to their needs.

\subsection{InfoInternet in schools}

Schools play an important role in social and symbolic integration. 
With InfoInternet the schools begin to play a new role, that of diffusion of technology and equitable access to information to children and transitively to their families.
Traditionally, school contributes to equalise social differences, offering children an equal baseline to start from in a society. 
\textbf{In rural communities} with high levels of poverty, the school is a fundamental space for reducing the economic divide. Various technologies that are not available in home or other community spaces, are often accessible through schools \cite{Sanchez2010IDC}. 
For example, in India, when starting school the student receives a tablet, thus ensuring that the student gets access to electronic information.
\jj{We need to get more details about this situation, who gives the tablets, the state, InfoInternet project? Which schools, the rich ones, all of them, even the ones in the poor communities? Could he direct us to references about this? \cj{The state gives, i.e. the school. We could look up the news articles, but I think it is enough with the one sentence we have now.}}

\subsubsection{The teachers role}
Though school is one of the main places providing access to Internet, the teacher does not seem to be a fundamental agent in transmitting ICT knowledge to children, being surpassed by other members of the communities \cite{Sanchez2010IDC}. However, teachers do act as \textit{gatekeepers}, generating conditions for children to learn in the school how to use the ICTs. 
We thus need to consider the \textit{role and position} the teacher could have in supporting the adoption of InfoInternet, both for children and pedagogical reasons. 

Many of the pedagogical theories on learning, motivation and socialising have a built-in philosophy on \textit{adaptive learning}. This is because the reality in schools shows notable differences between the learning needs of each individual, e.g., 
in the same school often there are different learning needs and capabilities between classes at the same level and even between individuals in the same class.
The role of the teacher is to be aware of these differences and to \textbf{adapt the content to the variety of levels and needs}. 
Moreover, teachers also know the importance of the cultural and social context in the learning process. In this perspective, as emphasised by sociologically oriented theories (such as Bernstein's theory of code \cite{sadovnik1995knowledge,bernstein2000pedagogy} and Berger and Luckmann's theory of primary and secondary socialisation \cite{berger1966theSocialConstruction}), 
teaching should take into account the student's language, values, way of conduct and cultural ballast. As such, a main requirement for developing apps and content for schools is to \textbf{bring along the teachers} and their expertise and to make sure that the content is flexible enough to support diversity and be relevant for the community. 

\textbf{Technical support and mentoring for teachers} is observed to result in better adoption of technology.
For a new Info\-Internet-based app or infrastructure deployment, one should first study in what measure the teachers see the InfoInternet as a positive tool for improving learning conditions, accessing resources for teaching, and for improving educational results.
An unfortunate example is the statement from Patrick Muinda from the Ministry of Education and Sports in Uganda \cite{internetsociety2017digitaldivide} informing how schools were equipped with all the technology necessary, but the computers were left unpacked as the teachers were lacking the digital skills and knowledge to take them in use in the classroom. The teachers did not want to embarrass themselves in front of the children. 
We observe that the more technical support or/and mentoring teachers get regarding the use of computers and InfoInternet as tools and resources for supporting the teaching activity 
the better the adoption. There is a wide range of support for teachers and administration as part of InfoInternet, as direct support for end users is the core of the InfoInternet principle. 
\jj{It will be good to get more information about this, about how InfoInternet supports the users, from Josef.}

\subsubsection{Influence of InfoInternet on learning results}

We consider, based on existing studies \cite{ciampa2014learning,crompton2017TheUse}, that once children and educators get access to relevant information through Internet, their motivation and skills will increase. 
Already only with the technological infrastructure in place one can see positive results in \textbf{cognitive and digital skills} development.
The evaluation of the ``One Laptop per child'' (OLPC) program \cite{Kraemer2009OLP} in poor communities in rural Peru \cite{cristia2012technology} 
offers proof in this respect,
even without any follow-up from the tutors or teachers. 
Other evaluations from use of OLPC in Tanzania show that the initial interest in digital technology drops if it is not supported by network access. This is where the InfoInternet solution is needed. 
Once the adoption and motivation is assured, as we can see from comprehensive studies \cite{hassler2016tablet,crompton2017TheUse}, mobile technologies can viably support children with their learning tasks and bring positive outcomes on learning achievements.

However, one should not start from the premise that technology by itself brings improvement in the quality of education, both when it comes to teachers and students.
Even if in the OLPC the laptops came loaded with 200 books, the children used them mainly for activities that had little effect on education, e.g.: word processing, calculator, games, music or recording sound and video \cite{cristia2012technology}. 
Another three years study in U.S.~schools~\cite{vigdor2014scaling} shows that the increased availability of computers at home, together with high-speed Internet, are associated with significantly lower math and reading test scores in the middle grades and less frequent computer use for homework. Computing devices could also \textbf{crowd out studying effort} by offering new forms of recreational activities. 

Even more, the introduction of broadband Internet could result in widening racial and socio-economic achievement gaps. 
The authors of the U.S. study interpret these findings by that home computer technology is put to more productive use in households with more effective parental monitoring, or in households where parents can serve as more effective instructors in the productive use of online resources.
Students might use the computers and Internet mainly for different recreational and entertainment purposes if not guided and monitored by both parents and teachers.

\subsubsection{Meaningful integration into curricula}
For computers/technology and Internet to be effective, they need to be \textit{linked to specific curricula goals and learning activities}. 
Access to Internet and computers is however a prerequisite for children and educators to be able to access learning resources, and for building the computer skills of the children, which are so much necessary in the global work market today.
The use of mobile technologies paired with appropriate pedagogical tools and methods, used meaningfully within teaching and learning environments, can decrease school drop-out, increase the interest for STEM (Science, Technology, Engineering, Mathematics) \cite{STEAM2015book,cominsky2017learning} subjects and careers, and improve general proficiency of children in STEM \cite{grant2015using}. 

\textbf{Other educational programs} and projects are necessary to complement the InfoInternet initiative. These should help with ICT knowledge building, creating of a curricula that integrates InfoInternet and ICT into the learning activities, and encourage and motivate both students and teachers to use computer technology as tools to support their school activities.

\subsection{InfoInternet for community supported education}

Education is not limited to the school areas and is not only the responsibility of the teachers; the entire community and the environment in which the children are growing up influences the development of a child.\cite{gestwicki2015home}  

\textbf{Family members} also play an important role; as these gain computer competence, they often support their children in the same direction. 
The study of \cite{Sanchez2010IDC} notices that even though parents have low education (or are even illiterate) their willingness to help their children with school tasks is considerable (88\%).
This is especially the case in poor communities where \textit{children often cannot attend school} \cite{Antle2017TheEtics} because, e.g.: 
\jjTODO{ We should prove this assumption through our own user studies. Does Josef already have reports about this situation from the areas where he already run pilot studies?}
\begin{itemize}
\item children are required to contribute to domestic chores, 
\item the school is not easily accessible because of the long distances, sometimes combined with adverse weather, 
\item the community does not even have a school building.
\end{itemize}

\textbf{Special needs} could be addressed by \textit{alternative educational applications} supporting remote and individual learning. Connectivity and/or Internet access is usually a prerequisite for such applications.

\textbf{Volunteer programs} like ``Computers Are Free for Everyone (CAFFE)'' in Bangladesh \cite{Ahmed2015CAFFE}, 
``Code Club'' in UK \cite{smith2014codeClub}, 
or ``L\ae{}r Kidsa Koding'' in Norway, are examples of alternatives that can \textit{supplement the institutionalised education}. 
Such initiatives could foster local job creation and ICT related entrepreneurship for youth and women \cite{margolis2003unlocking} and provide this vulnerable segment with job relevant and lifelong learning skills and new professional and life opportunities.

\textbf{Through public locations}, e.g., community centres, market places, health centres, besides schools, InfoInternet intends to provide information for everyone in the community. 
The intention is to introduce InfoInternet progressively, so that knowledge and expertise gathered in one sector will contribute to quicker results and development in the other sectors.
\jjTODO{ We should prove this assumption through our own user studies}%
\jjTODO{ Does Josef have reports that have observed the same? \cj{Nice observations.}}%
It is an advantage to have several points of InfoInternet access and several kinds of people that are exposed to it (e.g., from health care, from education, from social care) because the number of users will also increase. Thus the entire community can contribute to knowledge sharing and support each other in the process of building digital skills. 
\jjTODO{ We should prove this assumption through our own user studies}%
When people get together in groups, like in centres, the user that learns quickest, will give the other users pointers and suggestions \cite{kumar2017usability}.

We have seen that aggregate content, including broadband content, on a village server can have positive effect on the whole local society, because it will:
\begin{itemize}
 \item Bring preferred content free of charge
 \item Add the community aspects to the digital layer of information, meaning that communities can contribute to the content, can vote on it, and encourage each other to participate at reaching learning goals. 
\end{itemize}
\jj{NOTE: This is based on Josefs comment: If this information is based on observation in the field, could we get more details from Josef about this?}

\textbf{For remote learning}, the InfoInternet provides local info hotspots, which are InfoInternet local technology centres situated in public spaces where children can go easier and at different times than at school. Such a \emph{local info hotspot} is meant to offer access to educational material for free, on the location, either through provided hardware like tablets and/or through a local wifi-spot for children to connect their own devices. The local info hotspot would also provide, in various forms, content from local storage, e.g., videos previously retrieved from the Internet would be stored and served on location. The educational material should be the classroom material, preferably adapted for remote learning needs. The local info hotspot functionalities are not putting significant burden on the mobile infrastructure, thus the free availability.
Maintenance of the local info hotspot is done by the community, thus the complexity and variety of the available technology and features is dictated by the existing relevant technical competence in that community. 

The example of maintenance has been demonstrated through the Gram Marg project in India, where the local IT person:
\begin{itemize}
 \item Takes the responsibility for the community
 \item Has the possibility to earn his daily living through providing Internet access and other services, such as online shopping. 
\end{itemize}
\jj{It will be good to get more information from Josef about this scenario.}

\jjTODO{Does InfoInternet have already established programs for developing such local competence (through workshops and courses), or does it plan to do that?}%
The hardware implementations can be of various kinds fitted to the specific field requirements of the respective area 
(e.g., energy through solar panels to power up wifi hardware, tablets, smartphones and other connected things).
Such concerns are studied in other projects (like DigI) which we build on.

\subsection{Facilitating other educational programs and initiatives}

InfoInternet is intended to reach the most remote areas, with scarcity of resources at all levels. 
People in these communities have only one to two dollars per month available for communication. The coast goal of two dollars is reachable through only buying cheap equipment like tablets, as it have been shown through the Gram Marg project. This financial restriction can also be overcame through sharing of devices in the community centres or schools.
In consequence, InfoInternet must not be tied to one single kind of device, but should be useful on anything with browser-like functionality (e.g., phones, laptops, PCs; often old models). 
InfoInternet would cater also for future developments, when sensor-based devices, like Internet of Things, will be taken in use to support learning. The local hotspots can serve as hubs for such a sensor-based communication. 
The introduction of InfoInternet will allow further digital education projects to come into the community. These are projects that rely on the existence of some kind of access to Internet. 

\textbf{Adapting existing applications} 
like CETA \cite{marichal2017ceta}, which is a mixed-reality environment, augmenting the Cuisenaire roads, a milestone in manipulatives for mathematics learning, and inspired by a popular commercial game called Osmo. The application was adapted to work with low-cost tablets delivered to schools in Uruguay through the OLPC program. 
This tangible interaction environment is especially interesting for InfoInternet as it focuses on using the limited features of the low-end tablets as an already existing technological infrastructure put in place by other programs.

\textbf{Sensor-based Platforms} as built in projects like \cite{putjorn2017designing}, funded by Thailand's government, require connectivity. This platform supports science learning among primary school children in underprivileged Northern Thailand.
This platform does not rely on traditional mobiles, tablets and desktop devices, but on devices that can be created out of low cost components  (e.g., created locally, as part of a ICT program in the school), and adapted for different school sciences. 

\textbf{Learning by Making} is another project \cite{cominsky2017learning}, funded by the U.S.~Department of Education, with the goal to develop a high-school curriculum aligned with the Next Generation Science Standards (NGSS) integrating all four aspects of STEM. The project has already developed a one year course approved as a college preparatory Science laboratory course. This is especially relevant for InfoInternet because the developed platform is optimised for use by students in high-needs rural communities with low Internet bandwidth. 

\textbf{Augmented Reality (AR)} finds novel applications in combining physical world with digital information when building learning platforms \cite{bacca2015mobile,huang2016animating,jevrabek2014specifics,kamarainen2013ecomobile,yilmaz2016educational}.
In a second stage of InfoInternet, when access to digital devices and digital skills will be more spread, we envision a digital education seemingly integrated in the local culture and environment. 
In this sense the applications and devices could also support learning outside the physical classroom, in the local environment. Integrating with, and basing the education on the natural environment, would support local community building and make children aware of the local values, making them capable to contribute to its protection and development. In this envisioned environment, InfoInternet will offer the possibility to put the local in an global context and augment the local based knowledge with information available elsewhere.

The children could be provided with a digital toolbox, consisting of different cheap electronic components put together locally at schools. These devices would contain sensors and technology that could involve all senses of the child and link the physical world and information browsing to provide contextual and location-specific information. 
Such applications have been explored in more developed countries like Finland \cite{alakarppa2017using}.

%% file: examples_submission.txt
\section{Applications on top of InfoInternet}\label{sec_examples}

We analyse the characteristics of InfoInternet-friendly educational applications correlated with the previously described 
(A) intended target group, 
(B) educational environment, and 
(C) the specifics of the InfoInternet technology. 
We are interested both in existing applications that can be adapted to InfoInternet, as well as future applications designed to work well on top of InfoInternet. 

\subsection{Content type}

\begin{description}
\item[Free and open content] is encouraged, as our users have low purchasing power.
The content producer/distributor could use Creative Commons licences. 
One can find a list of such existing applications in the 2016 the State of the Commons report\footnote{State of the Commons: \url{https://stateof.creativecommons.org/}}.

\item[Shareable content] is recommended so that it can be displayed on, and transferred between, mobile devices or computers in schools or at home. In the case of India, the state requires all educational material to be provided in electronic formats (e.g., pdf, epub) that can be taken by the children with their devices. 
\jj{Look for reference about the India case or ask Josef about reference.}
The content should also be thought to be accessible for people with low digital competence and low literacy. 
Mobile apps like SHAREit or Zapya,\footnote{\url{https://www-shareit.com/} and \url{http://www.izapya.com/}} allow to get content from a hotspot, carry it home, and then share it with other users locally.

\item[Allow adaptation and contribution] so that local communities can also create and adapt the content for their needs and in their own languages/dialects.
It is not enough to only translate into a language, but also make the content suitable and relevant for the social and cultural context in which it is introduced, for otherwise it will not be used.

\end{description}
One application that covers well these criteria
comes from the African Storybook Initiative\footnote{\url{http://www.africanstorybook.org/}} \cite{SaideAfricanStoryBook,AfricanStorybook2017Journal}, aiming to support and promote literacy in the languages of Africa, using digital storybooks distributed by means of web browsers and native mobile app.

\subsection{Education types}

Educational applications pertain to one of three main types of education: formal, informal and semi-formal education \cite{grant2015using}. 

\begin{description}
\item[Formal learning] is initiated, led and evaluated by a teacher or a trainer (schools and MOOCs are examples).
For this case, an application should be developed together with the teachers and fitted with the curricula.

\item[Informal learning] is happening unorganized and unplanned, and is being initiated by the individual itself. 
Participating in a hacker space, in community activities like Code Clubs, reading technical books at home, DIY (Do-it-Yourself) kits and activities, are examples.

\item[Semi-formal learning] shares characteristics with both formal and informal learning \cite{Eshach2007nonformal}.
The earlier mentioned African Storybook is semi-formal as the storybooks are created by educators to be used to improve the literacy of the many children in sub-Saharan Africa that are out of school or that reach forth grade without learning to read. However, these books are intended to be used without guidance, by the child alone or together with the parents, or also in schools and libraries. The project is working with teachers, librarians, and literacy development organisations in Eastern and Southern Africa to overcome the shortage of children books, particularly in local languages. 
\end{description}

\subsection{Technological limitations}

As a general guideline, applications should be created (or adapted) with restricted resources in mind. 
\begin{description}
\item[Scarcity of devices] or low-end and old hard/soft-ware versions \cite{Franquesa2017SPDC}. 
Devices sold in Global South emerging markets have mostly low-end specifications, 
characterized by slow computing power, small screen sizes with low resolution, and short battery life \cite{Sambasivan2017GoogleBillion}. 
Smartphones are falling significantly in price, but due to requirements of charging once per day have not reached the rural markets.
These are devices running outdated operating systems, due to longer life spans (at least five years), often bought second-hand upon being discarded from wealthier areas.
However, thanks to programs like OLPC and the development of cheap tablets and smartphones, these devices are more spread in schools.

The content delivered to these devices should not require high computing/processing power and should be suitable for sharing between different kinds of devices and follow the progressive enhancement principle \cite{Gustafson2015ProgrEnh}. 
In schools, a shared computer will still improve the learning activities if this can be used to retrieve information and learning resources from the Internet. The information could then be shared with the students through analogue means.

\item[Bandwidth restrictions] or low mobile data allowances.
Data costs can be a substantial proportion of monthly expenses, e.g., 5\% in Nigeria and 11\% in Uganda. According to an Internet.org report,\footnote{State of Connectivity Report 2015:\\ \url{https://fbnewsroomus.files.wordpress.com/2016/02/state-of-connectivity-2015-2016-02-21-final.pdf} } in order to reach half the world, an app's monthly mobile data consumption should be no more than 250MB  \cite{Sambasivan2017GoogleBillion}.

There is increased interest among the web community in technologies related to creating apps that work over unstable or poor connections \cite{firtman2016high}. 
Today's web technologies (e.g., Service Workers or Progressive Web Apps \cite{ater2017building}) allow for building such applications, but require sometimes extra resources, which companies are not willing to invest if they do not see the revenue. 
Though there are companies that are interested in expending/invest in emergent markets, most of these are interested in engaging these communities in the digital world as mare consumers/users. Supporting consumerism is not what it is desired for a healthy society.

Since InfoInternet aims at providing access to information, it fulfils the requirements of low data amounts, and opens for a business model where aided/subsidized usage will pay for the usage of low content. It also supports models where the government puts constrains on those using ISP or mobile licences on non-profit content, being held as education, agriculture and e-gov services.

\item[Off-line use] and local storage are an advantage, as in the African Storybook where users need not be online to read the stories; they are able to download stories on their device, or print them. It is also possible to create or adapt a story offline, and then upload it when Internet is available.

\end{description}

\jjTODO{
We are interested in \textit{the most necessary applications} that one would like to make accessible for the poor communities, and develop an extended version of these web/native apps that will have support for offline activities or poor Internet connections. Such applications might even be available and need to be identified. 
}

\subsection{Intended use}

Besides technical aspects, the design of the applications should also consider the social and cultural needs of the users. Thus, prior to technical developments, user research needs to be conducted in the regions of interest to know the existing social, cultural, political, and religious norms, and how these influence the applications. 
These will help to make the applications relevant for the needs of the respective community.

\begin{description}
\item[Users] should be well identified. 
Types of users relevant for educational applications include: teacher, child alone, a group of children (collaborating and communicating), parent, someone trained in a community centre.

\item[Context] is a defining factor. Identify whether the application is to be used in schools, library, at home, on the play-ground, in a community centre.
\textit{In a community/playground setting} collaboration and group activities should be leveraged. 
The authors of \cite{Robinson2017ShareDevices} have created a framework to help with sharing of technological resources. The framework \textit{Better Together} splits core components of a service onto separate phones to support richer mobile interactions. For example, some devices have particularly good cameras, while others offer a large storage capacity or a fast Internet connection. The toolkit is open source and can be extended, adopted to other needs.
\textit{In an out-of-school context} the content should support individual and remote learning. 

\end{description}

%% file: conclusion_submission.txt
\section{Conclusions and Further work}\label{sec_conclusion}

About half of the population of the Earth is not connected to the Internet. Thus, they suffer from the digital divide. When Internet enters a market it is mainly provided through mobile networks which are solely revenue oriented. Thus, content on the Internet is typically revenue generating content, such as movies. As a consequence non-profit content such as health, education, agriculture, or e-gov services are not promoted in the mobile Internet. InfoInternet promotes the free access to information under a sustainable business model, combining both of them with 2.5\% of the bandwidth for basic information, text, pictures and local video, whereas more then 97\% of the mobile network capacity can be used for commercial services.

This paper is intended to help those that build educational apps for InfoInternet (or similar), giving them guidelines for how to understand their users, their educational environment, and the technological challenges of the networking infrastructure. 
More details will be made available in an online technical report.
Our future work will provide also technical guidelines for developers and designers, and for content creators.

%% file: main_arXiv.bbl
\begin{thebibliography}{10}

\bibitem{Ahmed2015CAFFE}
N.~Ahmed, A.~M.~M. Haque, and L.~Doyle.
\newblock Entering the dream world of computers.
\newblock In {\em Proceedings of the Seventh International Conference on
  Information and Communication Technologies and Development}, ICTD '15, pages
  23:1--23:4. ACM, 2015.

\bibitem{alakarppa2017using}
I.~Alak\"{a}rpp\"{a}, E.~Jaakkola, J.~V\"{a}yrynen, and J.~H\"{a}kkil\"{a}.
\newblock Using nature elements in mobile ar for education with children.
\newblock In {\em Proceedings of the 19th International Conference on
  Human-Computer Interaction with Mobile Devices and Services}, MobileHCI '17,
  pages 41:1--41:13. ACM, 2017.

\bibitem{Anderson2009Freemium}
C.~Anderson.
\newblock {\em {Free: The Future of a Radical Price}}.
\newblock Hyperion Books, 2009.

\bibitem{Antle2017TheEtics}
A.~N. Antle.
\newblock The ethics of doing research with vulnerable population.
\newblock {\em ACM Interactions}, 24(6):74--77, Nov. 2017.

\bibitem{armoni2015scratch}
M.~Armoni, O.~Meerbaum-Salant, and M.~Ben-Ari.
\newblock From scratch to "real" programming.
\newblock {\em ACM Transactions on Computing Education (TOCE)},
  14(4):25:1--25:15, Feb. 2015.

\bibitem{ater2017building}
T.~Ater.
\newblock {\em {Building Progressive Web Apps: Bringing the Power of Native to
  the Browser}}.
\newblock O'Reilly Media, 2017.

\bibitem{Avgerou2008ict4d}
C.~Avgerou.
\newblock Information systems in developing countries: a critical research
  review.
\newblock {\em Journal of Information Technology}, 23(3):133--146, Sep 2008.

\bibitem{Azpiazu2017}
I.~M. Azpiazu, N.~Dragovic, M.~S. Pera, and J.~A. Fails.
\newblock Online searching and learning: Yum and other search tools for
  children and teachers.
\newblock {\em Information Retrieval Journal}, 20(5):524--545, Oct 2017.

\bibitem{bacca2015mobile}
J.~Bacca, S.~Baldiris, R.~Fabregat, S.~Graf, et~al.
\newblock {Mobile Augmented Reality in Vocational Education and Training}.
\newblock {\em Procedia Computer Science}, 75:49--58, 2015.

\bibitem{Benham2012misinformation}
A.~Benham, E.~Edwards, B.~Fractenberg, L.~Gordon-Murnane, C.~Hetherington,
  D.~A. Liptak, M.~Smith, C.~Thompson, and A.~P. Mintz.
\newblock {\em Web of Deceit: Misinformation and Manipulation in the Age of
  Social Media}.
\newblock Information Today, Inc., 2012.

\bibitem{berger1966theSocialConstruction}
P.~L. Berger and T.~Luckmann.
\newblock {\em The Social Construction of Reality: A treatise in the sociology
  of knowledge}.
\newblock Anchor Books, 1966.

\bibitem{bernstein2000pedagogy}
B.~B. Bernstein.
\newblock {\em Pedagogy, symbolic control, and identity: Theory, research,
  critique}.
\newblock Rowman \& Littlefield Publishers, 2nd edition, 2000.

\bibitem{Charmaz2014GTconstrBook}
K.~Charmaz.
\newblock {\em Constructing Grounded Theory}.
\newblock Introducing Qualitative Methods Series. SAGE Publications, 2nd
  edition, 2014.

\bibitem{ciampa2014learning}
K.~Ciampa.
\newblock Learning in a mobile age: an investigation of student motivation.
\newblock {\em Journal of Computer Assisted Learning}, 30(1):82--96, 2014.

\bibitem{cominsky2017learning}
L.~Cominsky, C.~Peruta, S.~Wandling, B.~McCarthy, and L.~Li.
\newblock {Learning by Making for STEM Success}.
\newblock In {\em Proceedings of the 2017 Conference on Interaction Design and
  Children}, IDC '17, pages 773--776. ACM, 2017.

\bibitem{groundingTheory2015book}
J.~Corbin and A.~Strauss.
\newblock {\em {Basics of Qualitative Research: Grounded Theory Procedures and
  Techniques}}.
\newblock SAGE Publications, 4th edition, 2015.

\bibitem{Corneliussen16LarKoding}
H.~G. Corneliussen and L.~Pr\o{}itz.
\newblock Kids code in a rural village in norway: could code clubs be a new
  arena for increasing girls digital interest and competence?
\newblock {\em Information, Communication \& Society}, 19(1):95--110, 2016.

\bibitem{cristia2012technology}
J.~Cristia, P.~Ibarrar{\'a}n, S.~Cueto, A.~Santiago, and E.~Sever{\'\i}n.
\newblock Technology and child development: Evidence from the one laptop per
  child program.
\newblock {\em American Economic Journal: Applied Economics}, 9(3):295--320,
  July 2017.

\bibitem{crompton2017TheUse}
H.~Crompton, D.~Burke, and K.~H. Gregory.
\newblock The use of mobile learning in pk-12 education: A systematic review.
\newblock {\em Computers and Education}, 110(Supplement C):51 -- 63, 2017.

\bibitem{derman2010anti}
L.~Derman-Sparks and J.~O. Edwards.
\newblock {\em Anti-bias education for young children and ourselves}.
\newblock National Association for the Education of Young Children, 2010.

\bibitem{Eshach2007nonformal}
H.~Eshach.
\newblock Bridging in-school and out-of-school learning: Formal, non-formal,
  and informal education.
\newblock {\em Journal of Science Education and Technology}, 16(2):171--190,
  2007.

\bibitem{firtman2016high}
M.~Firtman.
\newblock {\em {High Performance Mobile Web: Best Practices for Optimizing
  Mobile Web Apps}}.
\newblock O'Reilly Media, 2016.

\bibitem{Franquesa2017SPDC}
D.~Franquesa and L.~Navarro.
\newblock Sustainability and participation in the digital commons.
\newblock {\em ACM Interactions}, 24(3):66--69, Apr. 2017.

\bibitem{STEAM2015book}
X.~Ge, D.~Ifenthaler, and J.~M. Spector, editors.
\newblock {\em {Emerging Technologies for STEAM Education: Full STEAM Ahead}}.
\newblock Educational Communications and Technology: Issues and Innovations.
  Springer International Publishing, 2015.

\bibitem{gestwicki2015home}
C.~Gestwicki.
\newblock {\em Home, school, and community relations}.
\newblock Cengage Learning, 2015.

\bibitem{gordon2013beginnings}
A.~M. Gordon and K.~W. Browne.
\newblock {\em Beginnings \& beyond: Foundations in early childhood education}.
\newblock Cengage learning, 2013.

\bibitem{Graham2003ACMmisinfo}
L.~Graham and P.~T. Metaxas.
\newblock "of course it's true; i saw it on the internet!": Critical thinking
  in the internet era.
\newblock {\em Commun. ACM}, 46(5):70--75, May 2003.

\bibitem{grant2015using}
M.~M. Grant.
\newblock Using mobile devices to support formal, informal and semi-formal
  learning.
\newblock In {\em Emerging Technologies for STEAM Education}, pages 157--177.
  Springer, 2015.

\bibitem{Gustafson2015ProgrEnh}
A.~Gustafson.
\newblock {\em Adaptive web design: crafting rich experiences with progressive
  enhancement}.
\newblock New Riders Press, 2nd edition, 2015.

\bibitem{hassler2016tablet}
B.~Ha{\ss}ler, L.~Major, and S.~Hennessy.
\newblock Tablet use in schools: a critical review of the evidence for learning
  outcomes.
\newblock {\em Journal of Computer Assisted Learning}, 32(2):139--156, 2016.

\bibitem{Heeks2002ict4dBADs}
R.~Heeks.
\newblock {Information Systems and Developing Countries: Failure, Success, and
  Local Improvisations}.
\newblock {\em The Information Society}, 18(2):101--112, 2002.

\bibitem{Heeks2008ict4d}
R.~Heeks.
\newblock {ICT4D 2.0: The Next Phase of Applying ICT for International
  Development}.
\newblock {\em Computer}, 41(6):26--33, June 2008.

\bibitem{hilbert2012toward}
M.~Hilbert.
\newblock {Toward a conceptual framework for ICT for development: Lessons
  learned from the cube framework used in Latin America}.
\newblock {\em Information Technologies \& International Development},
  8(4):pp--243, 2012.

\bibitem{huang2016animating}
T.-C. Huang, C.-C. Chen, and Y.-W. Chou.
\newblock {Animating eco-education: To see, feel, and discover in an augmented
  reality-based experiential learning environment}.
\newblock {\em Computers \& Education}, 96:72--82, 2016.

\bibitem{jevrabek2014specifics}
T.~Je{\v{r}}{\'a}bek, V.~Rambousek, and R.~Wildov{\'a}.
\newblock Specifics of visual perception of the augmented reality in the
  context of education.
\newblock {\em Procedia-Social and Behavioral Sciences}, 159:598--604, 2014.

\bibitem{kamarainen2013ecomobile}
A.~M. Kamarainen, S.~Metcalf, T.~Grotzer, A.~Browne, D.~Mazzuca, M.~S.
  Tutwiler, and C.~Dede.
\newblock {EcoMOBILE: Integrating augmented reality and probeware with
  environmental education field trips}.
\newblock {\em Computers \& Education}, 68:545--556, 2013.

\bibitem{2015InternetReport}
M.~Kende, editor.
\newblock {\em Global Internet Report: Mobile Evolution and Development of the
  Internet}.
\newblock Internet Society, 2015.
\newblock (available at
  \url{https://www.internetsociety.org/globalinternetreport/2015/}).

\bibitem{kim2012comparative}
P.~Kim, E.~Buckner, H.~Kim, T.~Makany, N.~Taleja, and V.~Parikh.
\newblock A comparative analysis of a game-based mobile learning model in
  low-socioeconomic communities of india.
\newblock {\em International Journal of Educational Development},
  32(2):329--340, 2012.

\bibitem{Kraemer2009OLP}
K.~L. Kraemer, J.~Dedrick, and P.~Sharma.
\newblock One laptop per child: Vision vs. reality.
\newblock {\em Commun. ACM}, 52(6):66--73, June 2009.

\bibitem{Kumar2014misinfodetect}
K.~P.~K. Kumar and G.~Geethakumari.
\newblock Detecting misinformation in online social networks using cognitive
  psychology.
\newblock {\em Human-centric Computing and Information Sciences}, 4(1):14, Sep
  2014.

\bibitem{kumar2017usability}
N.~Kumar, N.~Karusala, A.~Seth, and B.~Patra.
\newblock Usability, tested?
\newblock {\em ACM Interactions}, 24(4):74--77, June 2017.

\bibitem{leu2015income}
D.~J. Leu, E.~Forzani, and C.~Kennedy.
\newblock Income inequality and the online reading gap.
\newblock {\em The Reading Teacher}, 68(6):422--427, 2015.

\bibitem{Madon2009DigitalInclusion}
S.~Madon, N.~Reinhard, D.~Roode, and G.~Walsham.
\newblock {Digital inclusion projects in developing countries: Processes of
  institutionalization}.
\newblock {\em Information Technology for Development}, 15(2):95--107, 2009.

\bibitem{maloney2010scratch}
J.~Maloney, M.~Resnick, N.~Rusk, B.~Silverman, and E.~Eastmond.
\newblock {The Scratch programming language and environment}.
\newblock {\em ACM Transactions on Computing Education (TOCE)}, 10(4):16, 2010.

\bibitem{margolis2003unlocking}
J.~Margolis and A.~Fisher.
\newblock {\em Unlocking the clubhouse: Women in computing}.
\newblock MIT press, 2003.

\bibitem{marichal2017ceta}
S.~Marichal, A.~Rosales, F.~G. Perilli, A.~C. Pires, E.~Bakala, G.~Sansone, and
  J.~Blat.
\newblock Ceta: Designing mixed-reality tangible interaction to enhance
  mathematical learning.
\newblock In {\em Proceedings of the 19th International Conference on
  Human-Computer Interaction with Mobile Devices and Services}, MobileHCI '17,
  pages 29:1--29:13. ACM, 2017.

\bibitem{martin2013here}
F.~Martin and J.~Ertzberger.
\newblock Here and now mobile learning: An experimental study on the use of
  mobile technology.
\newblock {\em Computers \& Education}, 68:76--85, 2013.

\bibitem{Medhi2011TOCHI}
I.~Medhi, S.~Patnaik, E.~Brunskill, S.~N. Gautama, W.~Thies, and K.~Toyama.
\newblock Designing mobile interfaces for novice and low-literacy users.
\newblock {\em ACM Trans. Comput.-Hum. Interact. (TOCHI)}, 18(1):2:1--2:28, May
  2011.

\bibitem{Mintz2002misinformation}
A.~P. Mintz, editor.
\newblock {\em Web of Deception: Misinformation on the Internet}.
\newblock Information Today, Inc., 2002.

\bibitem{mitlin2013urban}
D.~Mitlin and D.~Satterthwaite.
\newblock {\em {Urban Poverty in the Global South: Scale and nature}}.
\newblock Routledge, 2013.

\bibitem{morse2016GTcompile}
J.~M. Morse, P.~N. Stern, J.~Corbin, B.~Bowers, K.~Charmaz, and A.~E. Clarke.
\newblock {\em Developing grounded theory: The second generation}.
\newblock Routledge, 2016.

\bibitem{noll2018InfoInternet}
J.~Noll, S.~Dixit, D.~Radovanovic, M.~Morshedi, C.~Holst, and A.~S. Winkler.
\newblock {5G Network Slicing for Digital Inclusion}.
\newblock In {\em Proceedings of the 10th International Conference on
  COMmunication Systems \& NETworkS}, COMSNETS'18, (Jan.) Bangalore, India, Jan
  2018. IEEE.

\bibitem{Pegrum14mobilEdu}
M.~Pegrum.
\newblock {\em Mobile learning: Languages, literacies and cultures}.
\newblock Palgrave Macmillan, 2014.

\bibitem{PinterWXRC17idc}
A.~T. Pinter, P.~J. Wisniewski, H.~Xu, M.~B. Rosson, and J.~M. Carroll.
\newblock Adolescent online safety: Moving beyond formative evaluations to
  designing solutions for the future.
\newblock In P.~Blikstein and D.~Abrahamson, editors, {\em Conference on
  Interaction Design and Children (IDC)}, pages 352--357. {ACM}, 2017.

\bibitem{putjorn2017designing}
P.~Putjorn, P.~Siriaraya, C.~S. Ang, and F.~Deravi.
\newblock Designing a ubiquitous sensor-based platform to facilitate learning
  for young children in thailand.
\newblock In {\em Proceedings of the 19th International Conference on
  Human-Computer Interaction with Mobile Devices and Services}, MobileHCI '17,
  pages 30:1--30:13. ACM, 2017.

\bibitem{resnick2009scratch}
M.~Resnick, J.~Maloney, A.~Monroy-Hern{\'a}ndez, N.~Rusk, E.~Eastmond,
  K.~Brennan, A.~Millner, E.~Rosenbaum, J.~Silver, B.~Silverman, et~al.
\newblock Scratch: programming for all.
\newblock {\em Communications of the ACM}, 52(11):60--67, 2009.

\bibitem{Robinson2017ShareDevices}
S.~Robinson, J.~Pearson, M.~Jones, A.~Joshi, and S.~Ahire.
\newblock {Better Together: Disaggregating Mobile Services for Emergent Users}.
\newblock In {\em Proceedings of the 19th International Conference on
  Human-Computer Interaction with Mobile Devices and Services}, MobileHCI '17,
  pages 44:1--44:13. ACM, 2017.

\bibitem{sadovnik1995knowledge}
A.~R. Sadovnik.
\newblock {\em {Knowledge and Pedagogy: The sociology of Basil Bernstein}}.
\newblock Alex Publishing Corporation, 1995.

\bibitem{Sambasivan2017GoogleBillion}
N.~Sambasivan, N.~Jain, G.~Checkley, A.~Baki, and T.~Herr.
\newblock {A Framework for Technology Design for Emerging Markets}.
\newblock {\em ACM Interactions}, 24(3):70--73, Apr. 2017.

\bibitem{Sanchez2010IDC}
J.~S\'{a}nchez.
\newblock Digital inclusion in chilean in rural schools.
\newblock In {\em Proceedings of the 9th International Conference on
  Interaction Design and Children}, IDC '10, pages 364--367. ACM, 2010.

\bibitem{selwyn2013distrustingICT4Dedu}
N.~Selwyn.
\newblock {\em Distrusting educational technology: Critical questions for
  changing times}.
\newblock Routledge, 2013.

\bibitem{selwyn2013education}
N.~Selwyn.
\newblock {\em Education in a digital world: Global perspectives on technology
  and education}.
\newblock Routledge, 2013.

\bibitem{selwyn2016education}
N.~Selwyn.
\newblock {\em Education and technology: Key issues and debates}.
\newblock Bloomsbury Academic, 2nd edition, 2016.

\bibitem{Seufert2014Freemium}
E.~B. Seufert.
\newblock {\em {Freemium Economics: Leveraging Analytics and User Segmentation
  to Drive Revenue}}.
\newblock Morgan Kaufmann Publishers, 2014.

\bibitem{smith2014codeClub}
N.~Smith, C.~Sutcliffe, and L.~Sandvik.
\newblock {Code Club: Bringing Programming to UK Primary Schools Through
  Scratch}.
\newblock In {\em Proceedings of the 45th ACM Technical Symposium on Computer
  Science Education}, SIGCSE '14, pages 517--522. ACM, 2014.

\bibitem{internetsociety2017digitaldivide}
I.~Society.
\newblock Internet and education: Can this partnership close the digital
  divide?, 2017.
\newblock (Webinar:
  \url{https://www.internetsociety.org/events/community-forums/2017/q4/}).

\bibitem{AfricanStorybook2017Journal}
E.~Stranger-Johannessen and B.~Norton.
\newblock The african storybook and language teacher identity in digital times.
\newblock {\em The Modern Language Journal}, 101(S1):45--60, 2017.

\bibitem{Thornberg12GTinformed}
R.~Thornberg.
\newblock Informed grounded theory.
\newblock {\em Scandinavian Journal of Educational Research}, 56(3):243--259,
  2012.

\bibitem{trower2015creating}
J.~Trower and J.~Gray.
\newblock Creating new languages in blockly: Two case studies in media
  computation and robotics (abstract only).
\newblock In {\em Proceedings of the 46th ACM Technical Symposium on Computer
  Science Education}, SIGCSE '15, pages 677--677. ACM, 2015.

\bibitem{Unnikrishnan2016OfElephants}
R.~Unnikrishnan, N.~Amrita, A.~Muir, and B.~Rao.
\newblock Of elephants and nested loops: How to introduce computing to youth in
  rural india.
\newblock In {\em Proceedings of the The 15th International Conference on
  Interaction Design and Children}, IDC '16, pages 137--146, New York, NY, USA,
  2016. ACM.

\bibitem{unwin2009ict4d}
T.~Unwin, editor.
\newblock {\em {ICT4D: Information and Communication Technology for
  Development}}.
\newblock Cambridge University Press, 2009.

\bibitem{valk2010using}
J.-H. Valk, A.~T. Rashid, and L.~Elder.
\newblock Using mobile phones to improve educational outcomes: An analysis of
  evidence from asia.
\newblock {\em The International Review of Research in Open and Distributed
  Learning}, 11(1):117--140, 2010.

\bibitem{vigdor2014scaling}
J.~L. Vigdor, H.~F. Ladd, and E.~Martinez.
\newblock Scaling the digital divide: Home computer technology and student
  achievement.
\newblock {\em Economic Inquiry}, 52(3):1103--1119, 2014.

\bibitem{APC2016report}
G.~I.~S. Watch, editor.
\newblock {\em Economic, Social and Cultural Rights and the Internet}.
\newblock Association for Progressive Communications (APC) and International
  Development Research Centre (IDRC), 2016.
\newblock (available at
  \url{https://www.giswatch.org/sites/default/files/Giswatch2016_web.pdf}).

\bibitem{weintrop2017comparing}
D.~Weintrop and U.~Wilensky.
\newblock Comparing block-based and text-based programming in high school
  computer science classrooms.
\newblock {\em ACM Transactions on Computing Education (TOCE)}, 18(1):3, 2017.

\bibitem{SaideAfricanStoryBook}
T.~Welch, J.~Tembe, D.~Wepukhulu, J.~Baker, and B.~Norton.
\newblock The african storybook project: an interim report.
\newblock In {\em Language Rich Africa Policy Dialogue: The Cape Town Language
  and Development Conference}, pages 92--95. British Council, 2015.

\bibitem{Winschiers2017CommunityLab}
H.~Winschiers-Theophilus and A.~Peters.
\newblock Community lab, namibia university of science and technology.
\newblock {\em ACM Interactions}, 24(6):16--19, Nov. 2017.

\bibitem{yilmaz2016educational}
R.~M. Yilmaz.
\newblock Educational magic toys developed with augmented reality technology
  for early childhood education.
\newblock {\em Computers in Human Behavior}, 54:240--248, 2016.

\end{thebibliography}
